\documentclass[useAMS,usecolumn,usenatbib,usegraphicx]{mn2e}

\title[The Environments of Ly$\alpha$ Blobs]{Ly$\alpha$ Blobs Like Company : The Discovery of A Candidate 100 kpc Ly$\alpha$ Blob Near to A Radio Galaxy with A Giant Ly$\alpha$ halo, B3\,J2330+3927 at $z=3.1$\thanks{Based on data collected at Subaru Telescope, which is operated by the National Astronomical Observatory of Japan.}}
\author[Y. Matsuda et al.]{
\parbox[t]{\textwidth}{\vspace{-1cm}
Y.\ Matsuda,$^{\! 1}$\thanks{E-mail: yuichi.matsuda@durham.ac.uk} Y.\ Nakamura,$^{\! 2}$ N.\ Morimoto,$^{\! 2}$ Ian Smail,$^{\! 1}$ C.\ De Breuck,$^{\! 3}$ K. Ohta,$^{\! 4}$ T. Kodama,$^{\! 5}$ A. K. Inoue,$^{\! 6}$ T. Hayashino,$^{\! 7}$ K. Kousai,$^{\! 7}$ E. Nakamura,$^{\! 7}$ M. Horie, $^{\! 7}$  T. Yamada,$^{\! 2}$ M. Kitamura,$^{\! 2}$ T. Saito,$^{\! 8}$ Y. Taniguchi,$^{\! 8}$ I. Tanaka,$^{\! 9}$ and P. Hibon $^{\! 10}$}\\\\
$^{1}$ Department of Physics, Durham University, South Road, Durham, DH1 3LE, UK \\
$^{2}$ Astronomical Institute, Graduate School of Science, Tohoku University, Aramaki, Aoba-ku, Sendai 980-8578, Japan\\
$^{3}$ European Southern Observatory, Karl-Schwarzschild Strasse, 85748 Garching bei M\"{u}nchen, Germany\\
$^{4}$ Department of Astronomy, Kyoto University, Kyoto 606-8502, Japan\\
$^{5}$ National Astronomical Observatory of Japan, Mitaka, Tokyo 181-8588, Japan\\
$^{6}$ College of General Education, Osaka Sangyo University, 3-1-1 Nakagaito, Daito, Osaka 574-8530, Japan\\
$^{7}$ Research Center for Neutrino Science, Graduate School of Science, Tohoku Univsersity, Sendai 980-8578, Japan\\
$^{8}$ Research Center for Space and Cosmic Evolution, Ehime University, 2-5 Bunkyo-cho, Matsuyama 790-8577, Japan\\
$^{9}$ Subaru Telescope, National Astronomical Observatory of Japan, 650 North A'ohoku Place, Hilo, HI 96720, USA\\
$^{10}$ Korea Institute for Advanced Study, 207-43 Cheongnyagni 2-Dong, Dongdaemun-gu, 130-722, Korea}

\begin{document}

\date{Accepted ... ; Received ... ; in original form ...}

\pagerange{\pageref{firstpage}--\pageref{lastpage}} \pubyear{2009}

\maketitle

\label{firstpage}

\begin{abstract}

We present the discovery of a candidate of giant radio-quiet Ly$\alpha$ blob (RQLAB) in a large-scale structure around a high-redshift radio galaxy (HzRG) lying in a giant Ly$\alpha$ halo, B3\,J2330+3927 at redshift $z=3.087$. We obtained narrow- and broad-band imaging around B3\,J2330+3927 with Subaru/Suprime-Cam to search for Ly$\alpha$ emitters (LAEs) and absorbers (LAAs) at redshift $z=3.09\pm0.03$. We detected candidate 127 LAEs and 26 LAAs in the field of view of $31' \times 24'$ ($58 \times 44$ comoving Mpc). We found that B3\,J2330+3927 is surrounded by a 130 kpc Ly$\alpha$ halo and a large-scale ($\sim 60 \times 20$ comoving Mpc) filamentary structure. The large-scale structure contains one prominent local density peak with an overdensity of greater than 5, which is $8'$ ($15$ comoving Mpc) away from B3\,J2330+3927. In this peak, we discovered a candidate 100 kpc RQLAB. The existence of both types of Ly$\alpha$ nebulae in the same large-scale structure suggests that giant Ly$\alpha$ nebulae need special large-scale environments to form. On smaller scales, however, the location of B3\,J2330+3927 is not a significant local density peak in this structure, in contrast to the RQLAB. There are two possible interpretations of the difference of the local environments of these two Ly$\alpha$ nebulae. Firstly, RQLAB may need a prominent ($\delta \sim 5$) density peak of galaxies to form through intense star-bursts due to frequent galaxy interactions/mergers and/or continuous gas accretion in an overdense environment. On the other hand, Ly$\alpha$ halo around HzRG may not always need a prominent density peak to form if the surrounding Ly$\alpha$ halo is mainly powered by its radio and AGN activities. Alternatively, both RQLAB and Ly$\alpha$ halo around HzRG may need prominent density peaks to form but we could not completely trace the density of galaxies because we missed evolved and dusty galaxies in this survey.

\end{abstract}


\section{Introduction}

Ly$\alpha$ blobs (LABs) are large Ly$\alpha$ nebulae in the high redshift Universe. Since the 1980's, LABs have been discovered around high-redshift radio galaxies \citep[HzRGs,][]{1993ARA&A..31..639M, 2007A&ARv..15...67M}. As the size and luminosity of these radio-loud LABs (RLLAB) show correlation with their radio activities, the formation mechanisms of LABs are thought to be mainly related to their radio and AGN activities \citep{1997A&A...317..358V}. However, the formation mechanisms of the diffuse outer parts of the nebulae may be different from those of the bright central part related to AGN activities, since the outer parts of the nebulae show more quiescent kinematics than the central part and show more varied structures, such as filaments and bubbles \citep{2003MNRAS.346..273V,2003ApJ...592..755R}. Since the 2000's, LABs lacking strong radio sources or AGN have been discovered \citep[radio-quiet LAB, or RQLAB,][]{2000ApJ...532..170S}. RQLABs also show filamentary and bubble-like structures, and the morphology of the outer parts of RLLABs and RQLABs are very similar \citep{2004AJ....128..569M}. Although at least three possible ideas, such as cold gas accretion, galactic winds, and photoinonization by intense star-bursts or by obscured AGN, have been proposed to explain the formation mechanisms of the RQLABs \citep{2000ApJ...532L..13T, 2000ApJ...537L...5H, 2001ApJ...548L..17C}, there is no consensus yet \citep{2009ApJ...700....1G, 2009arXiv0902.2999D}. Both RLLABs and RQLABs often have bright sub-millimeter (sub-mm) sources and reside in overdense regions, they could be linked to massive galaxy formation in overdense environments \citep{2000ApJ...532..170S, 2003Natur.425..264S, 2003ApJ...583..551S, 2007A&A...461..823V}. It is possible that RLLABs and RQLABs are closely related objects, but this relation is still unclear.

While giant LABs are very rare objects, protoclusters and the surrounding large-scale structures often contain multiple giant LABs. The number densities of 100 kpc-scale LABs have been estimated to be $< 3 \times 10^{-7}$ Mpc$^{-3}$ from several blind surveys for LABs \citep{2006ApJ...648...54S, 2007MNRAS.378L..49S, 2009ApJ...693.1579Y}. Despite this rarity, the SSA22 protocluster at $z=3.09$ contains two $\sim 200$ kpc RQLABs with a spatial separation of $\sim 6$ comoving Mpc \citep{2000ApJ...532..170S}. The SSA22 protocluster is located at the central part of a filamentary large-scale structure with spatial extents of $\sim 60$ comoving Mpc \citep{2004AJ....128.2073H}. This structure also contains more than $\sim 30$ smaller ($30-80$ kpc) RQLABs \citep{2004AJ....128..569M}. The protocluster associated with a 250 kpc RLLAB, MRC\,1138--262 also contains a 80 kpc RQLAB with a spatial separation of $\sim 5$ comoving Mpc \citep{2004A&A...428..793K, 2007A&A...461..823V}.

What is the difference between RLLABs and RQLABs? Is there any difference between the local environments of RQLABs and RLLABs? Is it common that giant RL and RQLABs reside in the same structure? Until now, there has been no wide-field Ly$\alpha$ imaging observation around RLLABs comparable to the SSA22 protocluster survey. We selected one of the most well-studied HzRGs, B3\,J2330+3927 at $z=3.087$, which was known to have a Ly$\alpha$ nebulae with an extent of at least $6''$ from the previous long slit spectroscopy and thus is a possible candidate for a giant RLLAB \citep{2003A&A...401..911D}. B3\,J2330+3927 has possible evidence that it resides in an overdense environment of galaxies. The JCMT/SCUBA observations of B3\,J2330+3927 showed that it has at least two possible companion sub-mm sources lying within $\sim 2$ comoving Mpc \citep{2003Natur.425..264S}. Fortunately, the Ly$\alpha$ of B3\,J2330+3927 falls on almost the central wavelength of the $NB497$ filter, which was originally made for the SSA22 $z=3.09$ protocluster observations \citep{2004AJ....128.2073H}. Here we present results of our Ly$\alpha$ imaging observation around B3\,J2330+3927 with the $NB497$.

In this letter, we use AB magnitudes and adopt cosmological parameters, $\Omega_{\rm M} = 0.3$, $\Omega_{\Lambda} = 0.7$ and $H_0 = 70$ km s$^{-1}$ Mpc$^{-1}$. In this cosmology, the Universe at $z=3.1$ is 2.0 Gyr old and $1''.0$ corresponds to a physical length of 7.6 kpc at $z=3.1$.

\section[]{Observations and Data Reduction}

\begin{table}
\centering
\begin{minipage}{85mm}
  \caption{Summary of Observations }
  \begin{tabular}{@{}ccccc@{}}
  \hline
 Filter & $\lambda_{\rm cent}$/$\Delta \lambda$$^a$ & Exposure Time & $5 \sigma$ (lim)$^b$ & FWHM\\
  & (\AA/\AA) & (s) & (AB mag) & ($''$)\\
\hline
 $NB497$ & 4977/77 & 14400 ($1800 \times 8$) & 25.6 & 0.7-0.9 \\
 $B$ & 4417/807 & 2880  ($360 \times 8$) & 26.5 & 0.5-0.9 \\
 $V$ & 5447/935 & 4320  ($360 \times 12$) & 26.4 & 0.6-0.7 \\
 $BV$ & 4977/1742 & -- & 26.6 & 1.0 \\
\hline
\end{tabular}
$^a$The central wavelength and FWHM of the filters.\\
$^b$The $5 \sigma$ limiting magnitudes within $2''$ diameter apertures.\\
\end{minipage}
\end{table}

\begin{figure}
\centering
  \includegraphics[scale=0.36]{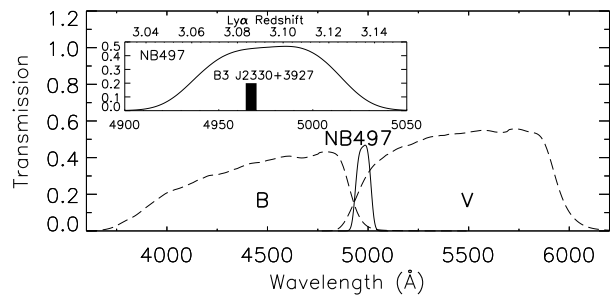}
  \caption{The transmission curves of the filters used for the observations. The solid line indicates the narrow-band filter, $NB497$. The dashed lines indicate broad-band filters, $B$ and $V$. The inset plot shows the zoomed $NB497$ transmission curve and the Ly$\alpha$ wavelength of B3\,J2330+3927 (the vertical thick bar). The profiles include the CCD quantum efficiency of Suprime-Cam, the transmittance of prime focus corrector, and the reflectivity of primary mirror of Subaru telescope.}
\end{figure}

\begin{figure*}
\centering
  \includegraphics[scale=.71]{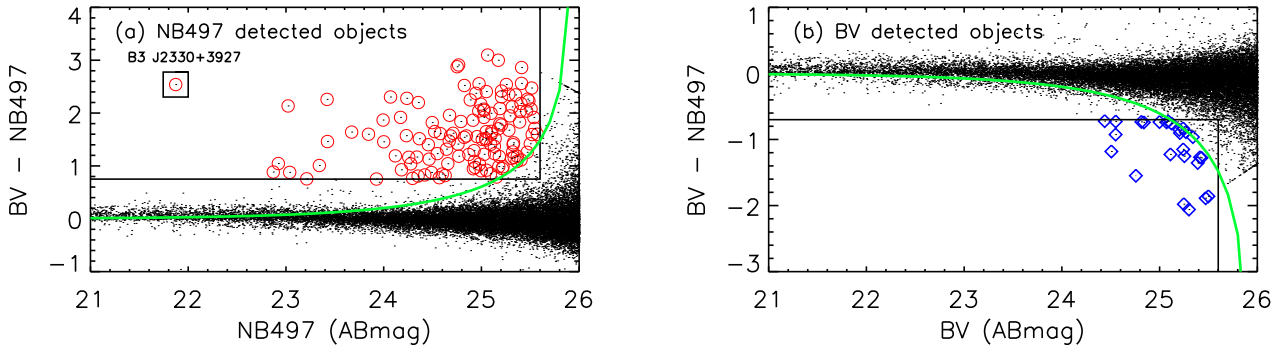}
  \caption{Colour magnitude plots. (a) $NB497$ vs $BV-NB497$ plot for $NB497$ detected sources. The solid line indicates colour and magnitude limits ($BV-NB497=0.75$ \& $NB497=25.6$). The red circles indicate candidate LAEs. B3\,J2330+3927 is the brightest LAE in this plot. (b) $BV$ vs $BV-NB497$ plot for $BV$ detected sources. The solid line indicates colour and magnitude limits ($BV-NB497=-0.7$ \& $BV=25.6$). The blue diamonds indicate candidate LAAs. The green curves show the 3.5 $\sigma$ uncertainty of $BV-NB497$ colour based on photometric errors of both $BV$ and $NB497$ for objects with constant $f_{\nu}$ spectra. All magnitudes and colours are measured with $2''$ diameter apertures.
}
\end{figure*}

We took narrow- and broad-band images centred at ($\alpha$,$\delta$) = 23:30:12.7, +39:30:41 (J2000.0) on 2007 November 09 (UT) with Suprime-Cam \citep{2002PASJ...54..833M} on the 8.2-m Subaru Telescope \citep{2004PASJ...56..381I}. Details of the observations are listed in Table~1. Suprime-Cam has a pixel scale of $0''.202$ and a field of view of $34' \times 27'$. The narrow-band filter, $NB497$, has a central wavelength of 4977 \AA\ and FWHM of 77 \AA\, which corresponds to the redshift range for Ly$\alpha$ at $z=3.062-3.126$. Fig.~1 shows the transmission curves of the $NB497$, $B$ and $V$-band filters, and the Ly$\alpha$ wavelength at the redshift of B3\,J2330+3927 ($z=3.087$).

 The raw data were reduced with {\sc sdfred20080620} \citep{2002AJ....123...66Y, 2003ApJ...582...60O} and {\sc iraf}. We flat fielded using the median sky image after masking objects. We did background sky subtraction adopting the mesh size parameter of 64 pixels ($13''$) before combining the images. We have confirmed that our results are not sensitive to these choices (see Section 3). Photometric calibration was obtained from the spectroscopic standard stars, LDS749B, and G191-B2B \citep{1990AJ.....99.1621O}. The magnitudes were corrected for Galactic extinction of $E(B-V)=0.124$ mag \citep{1998ApJ...500..525S}. The variation of the extinction in this field is small (peak to peak, $\pm 0.007$ mag) and thus it does not affect our results.

The combined images were aligned and smoothed with Gaussian kernels to match their seeing to a FWHM of $1''.0$. We made a $BV$ image [$BV=(2B+V)/3$] for the continuum at the same effective wavelength as $NB497$. The total size of the field analyzed here is $31'.2 \times 23'.6$ after removal of low S/N regions near the edges of the images. We also masked out the halos of the bright stars. The resultant total effective area is 699 arcmin$^2$ (corresponding to a comoving volume of $1.6 \times 10^5$ Mpc$^3$ at $z=3.1$). 

 Object detection and photometry were performed using {\sc SExtractor} version 2.5.0 \citep{1996A&AS..117..393B}. The object detections were made on the $NB497$ image (for Ly$\alpha$ emitters, or LAEs) and $BV$ image (for Ly$\alpha$ absorbers, or LAAs), using a Gaussian detection kernel with FWHM of 1$^{\prime\prime}$. We detected objects that had 5 connected pixels above $1.5 \sigma$ of the sky background rms noise. The magnitudes and colours are measured for each object in $2''$ diameter apertures.

\section[]{Results}

Fig.~2(a) shows a colour-magnitude plot for the $NB497$ detected objects. We selected 127 candidates as LAEs with the following criteria; (1) $NB497<25.6$ mag ($5~ \sigma$), (2) $BV - NB497 > 0.75$ mag ($EW_{\rm obs}>80$ \AA ),  (3) $\Sigma > 3.5$, where the $\Sigma$ is the ratio between the $NB497$ excess and the uncertainty of $BV-NB497$ colour based on photometric errors of both $BV$ and $NB497$ for objects with constant $f_{\nu}$ spectra. The Ly$\alpha$ luminosity limit of our LAE sample is $1.3 \times 10^{42}$ erg s$^{-1}$. We note that the contamination of [OII]$\lambda 3727$ emitters at $z=0.33$ in our LAE sample should be negligible thanks to the observed equivalent width limit of 80 \AA\ \citep[$<2\%$, e.g.,][]{2007ApJ...671..278G}.

We verified whether the entire field of view of the B3\,J2330+3927 field has overdensity compared with blank fields or not, using the $NB497$ image of Subaru-XMM Deep survey (SXDS) field taken by \citet{2004AJ....128.2073H}. As a result, the number density of LAEs in the B3\,J2330+3927 field is similar to that in SXDS at least for a sub-sample of bright LAE candidates with a large EW (L(Ly$\alpha$)$> 1.7 \times 10^{42}$ erg s$^{-1}$ and $EW_{\rm obs}> 120$ \AA ). Thus the B3\,J2330+3927 field does not show evidence for overdensity in the entire field of view. 

Fig.~2(b) shows a colour-magnitude plot for the $BV$ detected objects. We selected 26 candidates as LAAs with the following criteria; (1) $BV<25.6$ mag  ($12.5~ \sigma$), (2) $BV - NB497 < -0.7$ mag ($EW_{\rm obs}<-40$ \AA ), (3) $\Sigma > 3.5$, where the $\Sigma$ is the ratio between the $NB497$ depress and the uncertainty of $BV-NB497$ colour. We used the same equivalent width limit for LAAs as used in \citet{2000ApJ...532..170S}. They have spectroscopically confirmed that their LAA sample is at $z=3.1$ \citep{2000ApJ...532..170S}.

\begin{figure*}
  \includegraphics[scale=.65]{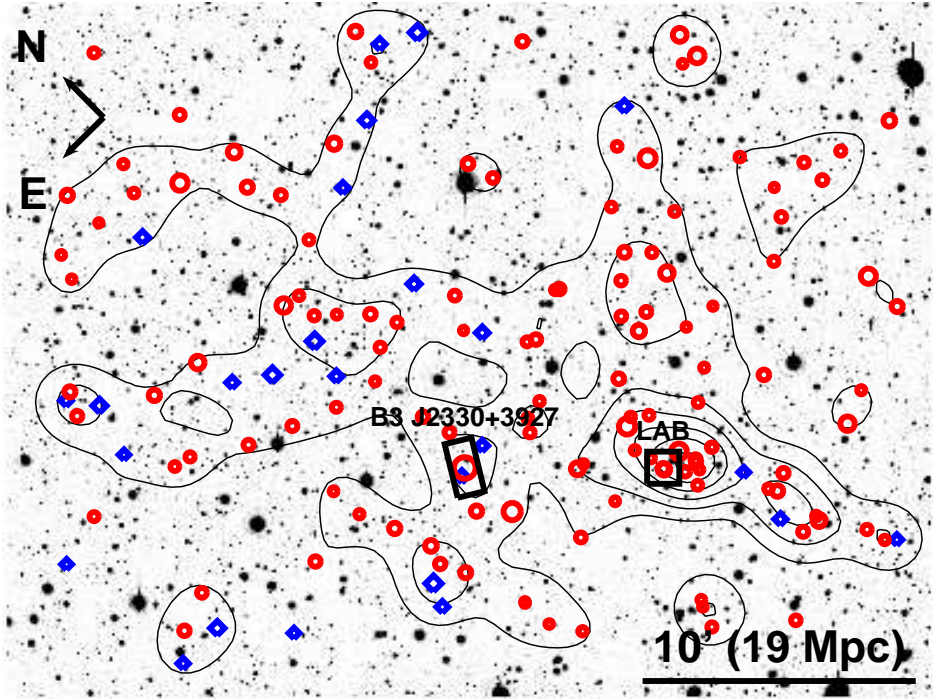}
  \caption{Sky distribution and smoothed density map of candidate LAEs and LAAs at $z\sim 3.09$ in the B3\,J2330+3927 field. The gray scale image is $NB497$ image. The thick bar shows the angular scale of $10'$ (19 comoving Mpc at $z=3.1$). The red circles and blue diamonds indicate candidate LAEs and LAAs, respectively. The size of the points are proportional to the luminosities (Ly$\alpha$ for LAEs and UV for LAAs). The rectangle indicates B3\,J2330+3927 and its major axis indicates the approximate direction of the radio axis. The square indicates the new RQLAB, LAB2330+3922. The surface density map of LAEs and LAAs are generated with a Gaussian smoothing kernel of $\sigma = 1.0'$, or FWHM $= 4.4$ comoving Mpc. The contours indicate deviations from the mean density of this field, $\delta\equiv(n-\bar{n})/\bar{n}=$0, 1, 2, 3, 4, and 5, equivalent to significance levels of 0, 1.2, 2.4, 3.6, 4.8, and 6.0 $\sigma$.}. 
\end{figure*}

Fig.~3 shows the spatial distribution of candidate 127 LAEs and 26 LAAs in this field and the smoothed density map of these objects. We note that neither LAE nor LAA is detected at the positions of two sub-mm sources around B3\,J2330+3927 in \citet{2003Natur.425..264S}. We made the density map smoothed with a Gaussian  kernel of $\sigma = 1'.0$, or FWHM $= 4.4$ comoving Mpc. The smoothing kernel size was chosen to match the median distance between the nearest neighbours in this sample.

We defined high-density regions (HDRs) as the regions with the overdensity $\delta\equiv(n-\bar{n})/\bar{n}>0$. We mark the position of B3\,J2330+3927 and the approximate direction of the radio axis \citep{2005MNRAS.363L..41P}. B3\,J2330+3927 is surrounded by a large HDR with an extent of $\sim 30' \times 10'$ ($\sim 60 \times 20$ comoving Mpc) . We have confirmed that it is difficult to reproduce such large HDRs from random distribution. We generated 10,000 density maps with the same size with 153 randomly distributed sources. The probability of finding HDRs with an area equal to or larger than that found around the B3\,J2330+3927 is less than $0.4\%$. Thus the large HDR around B3\,J2330+3927 should be a real large-scale structure. 

This large-scale structure contains one prominent density peak centred at ($\alpha$,$\delta$) = 23:29:53.5, +39:21:45 (J2000.0), which is $8'.0$ (15 comoving Mpc) south-west from B3\,J2330+3927. This peak has an overdensity of $\delta \sim 5.5$, which is greater than a significance level of $6~ \sigma$ compared with the average of the density fluctuation in this map. Moreover we discovered a candidate giant RQLAB lying very close to the density peak. However, the location of B3\,J2330+3927 is not a significant local density peak in this structure, in contrast to the new RQLAB. We note that there is no evidence for an overdensity of LAAs in the density peak, although LAEs appear to have a similar spatial distribution to LAAs on large scales in this field. We have confirmed that the results do not change significantly if we use different $NB497$ flat-field images taken by other projects and apply slightly different selection criteria for LAEs and LAAs.

\begin{table*}
 \centering
 \begin{minipage}{160mm}
  \caption{Properties of Ly$\alpha$ Blobs}
  \begin{tabular}{@{}ccccccccc@{}}
\hline
 ID & RA(J2000) & Dec(J2000) &  $\delta_{\rm gal}^a$ & Angular Size & EW$_{\rm obs}$ & log L(Ly$\alpha$) & $S_{1.4 GHz}^b$ & log M$_{\rm stel}^{c}$ \\
  & (h:s:m) & (d:s:m) & & ($''$/kpc) & (\AA ) & (cgs) & (mJy) & (M$_{\odot}$)\\
\hline
 LAB2330+3922  & 23:29:58.3 & +39:22:03 & $4.8\pm2.0$ & 13/100 & 356 & 43.6 & $<3$ & -- \\
 B3\,J2330+3927 & 23:30:24.9 & +39:27:11 & $1.5\pm1.3$ & 17/130 & 637 & 44.4 & $99.5\pm 3.0$ & $<11.94$\\
\hline
\end{tabular}
$^a$The local overdensities ($\delta\equiv(n-\bar{n})/\bar{n}$) derived from the smoothed density map of LAEs and LAAs. The uncertainties are calculated from the Poisson error of the number of objects within the Gaussian smoothing kernel at the positions.\\
$^b$The 1.4 GHz flux density from NRAO VLA Sky Survey \citep[NVSS,][]{1998AJ....115.1693C}\\
$^c$The stellar mass from \citet{2007ApJS..171..353S}\\
\end{minipage}
\end{table*}

\begin{figure}
\centering
  \includegraphics[scale=.44]{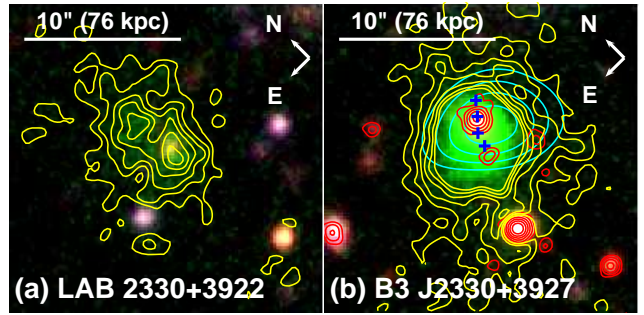}
  \caption{Colour images ($B$ for blue, $NB497$ for green, $V$ for red) of LABs. (a) A candidate RQLAB, LAB\,2330+3922. (b) B3\,J2330+3927. The size of the images are $20'' \times 20''$ ($152 \times 152$ kpc). The contours indicate Ly$\alpha$ surface brightness of 2, 4, 6, 8, 10, and 12 $\times 10^{-18}$ erg s$^{-1}$ cm$^{-2}$ arcsec$^{-2}$ (1.5, 3, 4.5, 6, 7.5, and 9 $\sigma$ per arcsec$^2$ in continuum subtracted $NB497$ image). The red contours indicate $K$-band source positions and the cyan contours indicate CO(4-3) emission \citep{2003A&A...401..911D}. The blue crosses indicate the radio knots from \citet{2005MNRAS.363L..41P}.}
\end{figure}

Fig.~4 shows $B, NB497, V$ colour images of the candidate RQLAB, LAB2330+3922 and B3\,J2330+3927. The RQLAB has an extent of $13''$ (100 kpc) and it is one of the largest LABs known to date. The RQLAB has two Ly$\alpha$ peaks and was initially detected as two different LAEs. B3\,J2330+3927 also has giant Ly$\alpha$ nebula with an extent of $17''$ (130 kpc). We listed the properties of these two LABs in Table~2. The Ly$\alpha$ luminosities and observed Ly$\alpha$ equivalent widths are measured using isophotal apertures with a threshold of $2 \times 10^{-18}$ erg s$^{-1}$ cm$^{-2}$ arcsec$^{-2}$ (the lowest contours in Fig.~4). We have confirmed that there is no other prominent LAB in this field except for these two LABs, searching for LABs with the detection threshold of $2 \times 10^{-18}$ erg s$^{-1}$ cm$^{-2}$ arcsec$^{-2}$ and the selection criteria of $BV-NB497\ge1.5$ mag, $NB497\le25$ mag, isophotal area $\ge 50$ arcsec$^2$. We have also confirmed that the results do not change significantly if we use different mesh sizes for sky-subtraction in the data reduction process.

\section[]{Discussion and Conclusions}

We discovered a new candidate 100 kpc RQLAB, which appears to lie in the same large-scale structure around the RLLAB, B3\,J2330+3927. This discovery is unlikely to be serendipitous. If we use the upper limit of the number density of 100 kpc-scale LABs from previous blind surveys for LABs at $z=2-5$  \citep{2006ApJ...648...54S, 2007MNRAS.378L..49S, 2009ApJ...693.1579Y}, the estimated probability to find a new 100 kpc LAB in the survey volume of our observation is only $< 5\%$. Thus it is more likely that this large-scale structure is a special environment for LABs to form. However, we need spectroscopic redshifts of the new RQLAB and the large-scale structure to investigate whether the new RQLAB and the RLLAB are really in the same structure or not.

On smaller scales, the new RQLAB appears to lie in the local density peak in the large-scale structure, while the RLLAB, B3J2330+3927 does not. There are two possible interpretations of the difference of the local environments of these two LABs. Firstly, RQLAB may need a prominent ($\delta \sim 5$) density peak of galaxies to form through intense star-bursts due to frequent galaxy interactions/mergers and/or continuous gas accretion in an overdense environment. The prototypes of RQLABs, SSA22 LAB1 and LAB2 are also known to reside in overdensities of star-forming galaxies of greater than $\delta \sim 5$ \citep{2000ApJ...532..170S}. On the other hand, RLLAB may not always need a prominent density peak to form if it is mainly powered by its radio and AGN activities. Alternatively, both RQLAB and RLLAB may need prominent density peaks to form but we could not completely trace the density of galaxies because we missed evolved and dusty galaxies in our survey. B3\,J2330+3927 has two possible sub-mm companions although neither has a LAE/LAA counterpart \citep{2003Natur.425..264S}. Thus it is still possible that B3\,J2330+3927 is associated with an overdense region of evolved and dusty galaxies. Future deep, wide-field NIR and sub-mm observations will be useful to investigate these possibilities by providing a complete map including evolved and dusty galaxies in these environments. 

\section*{Acknowledgments}

We thank the referee, Bram Venemans for careful reading the manuscript. We also thank Dave Alexander, Mark Swinbank, Jim Geach, Jim Mullaney, Richard Bower and Rob Ivison for help and useful discussions. YM and IRS acknowledge support from STFC.


\label{lastpage}

\end{document}